\documentclass[useAMS,usenatbib,usegraphicx]{mn2e}
\usepackage{color}

\title[Deuterium fractionation in the protosolar nebula]{Chemical fractionation of deuterium in the protosolar nebula}
\author[J. Kalv\=ans et al.]{
J. Kalv\=ans$^{1}$\thanks{E-mail: juris.kalvans@venta.lv},
I. Shmeld$^{1}$,
J. R. Kalnin$^{1}$,
S. Hocuk$^{2}$\\
$^{1}$Engineering Research Institute "Ventspils International Radio Astronomy Centre" of Ventspils University College,\\
In$\check{z}$enieru 101, Ventspils, LV-3601, Latvia\\
$^{2}$Max Planck Institute for Extraterrestrial Physics,\\
Giessenbachstrasse 1, 85748 Garching, Germany
}
\begin{document}

\date{Accepted 201X Month X. Received 201X Month X; in original form 2015 May 14}

\pagerange{\pageref{firstpage}--\pageref{lastpage}} \pubyear{201X}

\maketitle

\label{firstpage}

\begin{abstract}
Understanding gas-grain chemistry of deuterium in star-forming objects may help to explain their history and present state. We aim to clarify how processes in ices affect the deuterium fractionation. In this regard, we investigate a Solar-mass protostellar envelope using an astrochemical rate-equation model that considers bulk-ice chemistry. The results show a general agreement with the molecular D/H abundance ratios observed in low-mass protostars. The simultaneous processes of ice accumulation and rapid synthesis of HD on grain surfaces in the prestellar core hampers the deuteration of icy species. The observed very high D/H ratios exceeding 10 per cent, i.e., super-deuteration, are reproduced for formaldehyde and dimethyl ether, but not for other species in the protostellar envelope phase. Chemical transformations in bulk ice lower D/H ratios of icy species and do not help explaining the super-deuteration. In the protostellar phase, the D$_2$O/HDO abundance ratio was calculated to be higher than the HDO/H$_2$O ratio owing to gas-phase chemistry. Species that undergo evaporation from ices have high molecular D/H ratio and a high gas-phase abundance.
\end{abstract}


\section{Introduction}
\label{intro}

Deuterium fractionation $R_{\rm D}$ or D/H\,=\,[XD]/[XH] represents the enhancement of D content in molecules (the notation [X] stands for the relative abundance of chemical species X). $R_{\rm D}$ for molecules in Solar-system bodies and star-forming regions is often found to be much higher than the cosmic D/H abundance ratio of $\approx10^{-5}$. To understand chemical processes in prestellar cores, protostellar envelopes, and protoplanetary discs, it is crucial to `decipher' the information contained in the measured $R_{\rm D}$ values of molecules in different objects.

Deuterium fractionation in star-forming regions has recently become an increasingly active research field due to important observational, experimental, and theoretical advances. Overviews are available in, e.g., \cite{Albertsson13}, \cite{Taquet14}, \cite{Awad14}, and \citet{Ceccarelli14}. Specific attention has been paid to deuterium fractionation of water by \citet{Taquet13a}, \cite{Furuya13}, \cite{Albertsson14}, \cite{Dishoeck13}, and \citet{Wakelam14}.

The modeling work by \citet{Taquet13a} and \citet{Kalvans13} indicates that deuterium-rich ice molecules mostly are concentrated on the outer surface of ice mantles residing on interstellar or circumstellar grains. To put it another way, ice molecules that are formed early in the evolution of a contracting cloud core (e.g., H$_2$O) typically have low $R_{\rm D}$, while species that are synthesized in the later stages of prestellar cores typically, and are abundant on the outer surface, have a high D content (e.g., CH$_3$OH). Moreover, the molecules in bulk ice are chemically uncoupled from the chemical processes on the surface and in the gas. The average $R_{\rm D}$ for bulk molecules is substantially lower than previously thought.

The different deuteration of surface and bulk-ice molecules can only be reproduced with astrochemical models that consider bulk-ice species as a separate phase. Recent deuterium chemistry models that consider this are those of \citet{Taquet14} and \citet{Furuya15}. The former considers circumstellar envelopes in a physically detailed way and confirms that D/H is high for surface and low for bulk ice species. The monolayer structure of this model \citep{Taquet12a} allowed these authors determining the detailed composition of ice as a function of ice depth, measured in monolayers.

The above papers consider the icy mantles as rigid structures, where molecule diffusion and reactions occur only in the surface monolayer. However, photoprocessing of subsurface ice can be a major route for chemical synthesis. With the help of numerical simulations, such processes have been studied by, e.g., \citet{Kalvans10,Garrod13}, and \citet{Chang14}. Chemical processing of bulk ice may introduce changes in the D/H ratio for icy species \citep{Kalvans13}. Therefore, the specific aim of the present study is to investigate the deuterium fractionation of circumstellar molecules by using a model that considers active subsurface ice chemistry. For this purpose, a model was employed that considers time-dependent physical conditions in an infalling protostellar envelope (Sect.\,\ref{mphys}), ice depth-dependent composition of mantle layers on grains, and an extensive reaction network of deuterated molecules (Sect.\,\ref{nchem}). This model adds to the understanding of the processes that govern deuterium fractionation in gas, surface, and mantle in general (Sect.\,\ref{r-dice}), and for specific compounds (Sect.\,\ref{rspec}). The outcome of these studies is summarized in Sect.\,\ref{dconcl}.

\section{Methods}
\label{meth}

Our numerical simulation is based on the model `Alchemic-Venta' \citep{Kalvans15c,Kalvans15b}. A short description of the model is presented below; it includes a detailed description for items that differ from \citet{Kalvans15c}.

\subsection{Physical model}
\label{mphys}
%
\begin{figure}
 \vspace{-2.5cm}
  \hspace{-1.6cm}
  \includegraphics[width=18.0cm]{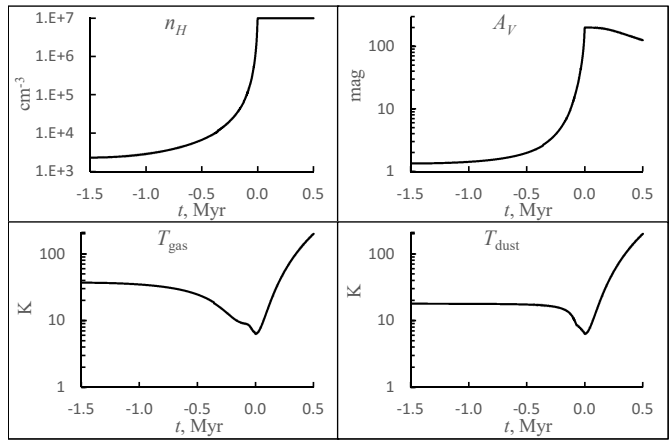}
 \vspace{-17cm}
 \caption{Evolution of temperature, density, and interstellar extinction. Density, gas temperature, and dust temperature are plotted for the modeled center part of the cloud core. $A_{\rm V}$ corresponds for a line of sight passing through the center. The protostar is formed at time $t=0$.}
 \label{att-phys}
\end{figure}
%
\begin{table}
 \centering
 \begin{minipage}{80mm}
\caption{Density at the center of the cloud core, radius $r_1$ of the core, and core mass at the start, at starbirth time, and at the end of the simulation.}
\label{tab-phys}
  \begin{tabular}{@{}lccc@{}}
  \hline
$t$, Myr & $n_{\rm H}$, cm$^{-3}$ & $r_1$, AU & $M_{\rm core}$, $\rm M_\odot$ \\
 \hline
-1.5 & $2.3\times10^3$ & $1.0\times10^4$ & 0.035 \\
0 & $1.0\times10^7$ & $1.0\times10^4$ & 1.18 \\
0.5 & $1.0\times10^7$ & $1.0\times10^3$ & --- \\
\hline
\end{tabular}\\
\end{minipage}
\end{table}
Following the approach of \citet{Garrod06}, we adopt a two-stage hot-core model in order to represent a contracting prestellar molecular cloud core (Phase~1, integration time $t\leq0$) and the subsequent heating of a circumstellar nebula (Phase~2, $t>0$). With this we assume that the protostar is formed at $t=0$.

The simulation is performed for a single point (0D model) at the centre of a spherical core with radius $r_1$. The core was assumed to have a Plummer-like density profile with radius $r_0=1.3\times10^5$\,AU for the central density plateau and power-law parameter $\eta=5.0$. The core is embedded in a surrounding cloud with a fixed column density $N_{\rm out}=1\times10^{21}$\,cm$^{-2}$. In Phase~1, the core undergoes free-fall gravitational collapse from an initial central density of $n_{\rm H,0}=2.3\times10^3$\,cm$^{-3}$, which increases to $10^7$\,cm$^{-3}$ over a time-scale of 1.5\,Myr. The central plateau radius $r_0$ decreases proportionally to $n_{\rm H}^{-1/2}$ \citep{Keto10,Taquet14} until it reaches $2\times10^3$\,AU, representing a centrally highly concentrated core.

These quantities were used to calculate the total column density $N_{\rm H}$ in the center of the spherical cloud core according to the Eq.\,(1) of \citet{Kalvans15c}. The obtained value of $N_{\rm H}$ is then used to calculate the interstellar extinction $A_{\rm V}$\,(mag), for a line of sight through the center of the core. This $A_{\rm V}$ value is twice as high as that from cloud edge to its center. We use a conventional $N_{\rm H}/A_{\rm V}$ ratio of $2.0\times10^{21}$ \citep[see][and references therein]{Valencic15}. For calculating $A_{\rm V}$, we consider $N_{\rm out}$ as well as the matter inside $r_1$. Our simulation starts with $A_{\rm V}=1.34$\,mag, corresponding to about 0.67\,mag in the simulation of \citet{Furuya15}.

For the cold core phase, it was assumed that the cloud is externally heated, i.e., its temperature is governed by interstellar extinction. Dust temperature $T_{\rm dust}$ was calculated as a function of $A_{\rm V}$ with the method outlined by \citet{Garrod11}. The relation between gas temperature $T_{\rm gas}$ and $A_{\rm V}$ was derived from the data of the 3D collapsing cloud models by \citet{Hocuk16}. The difference between the two temperatures becomes less than 0.5\,K when $A_{\rm V}>17$\,mag and they equalize when $A_{\rm V}>89$\,mag (remember that with $A_{\rm V}$ here we mean extinction along the whole line of sight).

In the second stage, the modeled gas parcel undergoes heating up to 200\,K by energy influx from the newly formed protostar according to the T2 profile of \citet{Garrod06}. The warm-up time-scale is related to the lifetime of the dense envelope around the protostar. Infalling envelopes are thought to exist around protostars of Classes 0 and I, and around objects with flat spectral energy distribution \citep{Stahler05,Hartmann09}. \citet{Evans09} estimate that the median lifetimes for these objects are 0.1--0.16\,Myr, 0.44--0.54\,Myr, and 0.35--0.40\,Myr, respectively. \citet{Maury11} estimate a median lifetime of 0.04--0.09\,Myr for Class\,0 objects. This means that the combined median lifetime of the envelope could lie in the range between 0.83 and 1.10\,Myr although there are significant uncertainties \citep{Schnee12,Carney16}. The observations, which we use for comparing results (Table\,\ref{tab-obs}), have targeted both Class\,0 and Class\,I objects.

For the purpose of the present model, we adopt a warm-up time-scale of 0.5\,Myr, approximately half of the entire lifetime of a typical collapsing envelope. Volatile icy species, such as CO and CH$_4$, evaporate near the end of the Class\,0 phase, and the icy mantles are completely evaporated 0.4\,Myr after starbirth, well into the Class\,I phase.

The gas parcel in consideration was assumed to be fully shielded from the protostellar radiation. Meanwhile, it was assumed that the radius of the core linearly decreased during the warm-up phase from the cloud core value of $r_1=10^4$\,AU to $10^3$\,AU, which is closer to the characteristic size of a protostellar nebula \citep[e.g.][]{Visser11}. Such a simple isopycnic (constant density) contraction was introduced to account for the mass loss in the circumstellar envelope, as it falls into the star and the protoplanetary disc. The resulting change in $A_{\rm V}$ has nearly no effect on the results. Fig.\,\ref{att-phys} shows the evolution of physical conditions in the modeled cloud core.

Table\,\ref{tab-phys} summarizes the macrophysical parameters at key moments of the simulation. Because $n_{\rm H}$, $r_1$, and $r_0$ vary with time, the core mass $M_{\rm core}$ encompassed within $r_1$ is not constant.

\subsection{Chemical network}
\label{nchem}

The model adopts the publicly available full deuterium reaction network from \citet{Albertsson13}. This network was generated from the osu.2009 reaction database and does not discern between H and D atoms that are attached to different heavy atoms in a single molecule. In other words, H and D are interchangeable as they freely undergo intramolecular diffusion in reactions. For example, the cosmic-ray-induced photodissociation of the hydroxymethyl radical $\rm CH_2OD$  is included in the database as two reactions:
   \begin{equation}
   \label{chem1}
\mathrm{CH_2OD} + h\nu \longrightarrow \mathrm{CH_2} + \mathrm{OD},
   \end{equation}
\[
\mathrm{CH_2OD} + h\nu \longrightarrow \mathrm{CHD} + \mathrm{OH}.
\]
The second product set requires an intramolecular exchange of H and D atoms before the molecule is split in two. This behavior is not always possible, especially, when low energies and large molecules are involved. However, in the reaction set, intramolecular mixing of H and D is assumed to occur for all reactions, regardless of their type, reactants and products.

The latter aspect decreases the extent of selective deuterium fractionation. For example, let us assume that CH$_2$OH is produced via CH$_2$+OH surface reaction. If CH$_2$ is more enriched in deuterium than OH, then this information is lost in the synthesis, as well as when the molecule disintegrates via reactions (\ref{chem1}). This means that $R_{\rm D}$ for the C--H bond is lowered and that of the O--H bond is increased by fictitious intramolecular diffusion. However, H and D are not fully interchangeable in the interstellar medium. This is manifested by the low CH$_3$OD/CH$_2$DOH abundance ratio, $\approx$10 per cent, as observed by \citet{Parise06}. With H and D interchangeable, such a low ratio cannot be reproduced.

To change this unsatisfying situation and achieve more realistic results in abundance ratios of complex organic molecule (COM) isotopologues, we revised the reaction network by hand. The products that involve intramolecular diffusion of H and D were removed for reactions involving CH$_2$OH, CH$_3$OH, HCOOCH$_3$, HCOOH, CH$_3$CHO, CH$_3$OH$^+$, CH$_3$OH$_2^+$, CH$_3$O(H)CH$_3^+$, and CH$_3$CHO$^+$. For the methyl formate cation, the species COOCH$_4^+$ were replaced with HCOOCH$_3^+$. We also add methoxy radical CH$_3$O to the network. For the latter, necessary reaction data was adopted from the network employed by \citet{Garrod08}. These changes do not fully eliminate the interchangeability for H and D in COMs.

Several additional changes were introduced in this network. The desorption energy $E_D$ for H$_2$ was adopted to be 430\,K \citep{Cazaux04, Heine04, Garrod06}. For HD and D$_2$, the values were assumed to be higher by 2 and 32\,K, respectively \citep{Kristensen11}. $E_D$ was taken to be 450\,K for H and 471\,K for D \citep{Caselli02,Aikawa12}. The network was updated with reactions 5--12, 14, and 16--19 from \citet{Kalvans15c}. The deuterium analogues for these reactions were included, with an exception of $\rm D+CO\longrightarrow DCO$. The activation barrier of this reaction (1400\,K) in the network of \citet{Albertsson13} is already lower than the 1600\,K assumed by \citet{Kalvans15c}.

Another parameter, which may affect $R_{\rm D}$ of interstellar and circumstellar molecules, is the ortho/para ratio of the H$_2$ molecule \citep{Walmsley04,Flower04,Flower06-1,Flower06-2}. Molecular hydrogen formed on the grains is released into the gas phase with the statistical ortho/para ratio of 3:1. The subsequent interaction with H$^+$ then reduces this value towards thermal equilibrium value \citep{Pagani09,Sipila13,Sipila15}. This happens even before the formation of the dense core. In the present study, we do not consider the spin states of H$_2$ or other species. This approach eases calculations of the nearly 80,000 molecular processes in the model and can be justified by the following discussion.

First, ice formation in the present model occurs in a dense medium, where the H$_2$ ortho/para ratio is assumed to be low, and therefore has a limited practical effect. More than 90 per cent of ice molecules are deposited when $A_{\rm V}>4$\,mag and ortho/para ratio is around $10^{-3}$, according to the models of \citet{Sipila13} and \citet{Furuya15}.

Second, the main source for a high abundance ratio of gas-phase neutral D atoms with respect to similar H atoms (atomic D/H) in the prestellar Phase~1 is HD photodissociation, as in \citet{Furuya15}. The photodissociation of HD occurs faster than that of H$_2$ due to shielding effects. As a result, the atomic D/H in the modeled (translucent) core is higher than the cosmic D/H ratio and largely independent of binary gas-phase reactions involving the isotopologues of H$_3^+$, where the H$_2$ ortho/para ratio is important. The latter aspect means that H$_2$ spin states do not have a decisive effect on ice deuteration in diffuse medium either.

A third point regarding the effect of H$_2$ spin states on ice deuteration is that there is considerable evidence that the ortho/para ratio decreases when H$_2$ interacts with interstellar ice analogues \citep{Lebourlot00,Watanabe10,Sugimoto11,Chehrouri11,Hama12}. As far as we know, this effect has not yet been considered in astrochemical models. Thanks to such interactions, the relaxation time of the H$_2$ ortho/para ratio in cold cores is reduced, reducing also the spatial and temporal extent of zones with high ortho/para ratio. This adds to the uncertainties regarding the H$_2$ ortho/para ratio, which limits the usefulness of the inclusion of H$_2$ spin states in the present study.

\subsection{Chemical model}
\label{mchem}
\begin{table}
 \centering
 \begin{minipage}{80mm}
\caption{Initial abundances of chemical species with respect to H nuclei.}
\label{tab-el}
  \begin{tabular}{@{}lc@{}}
  \hline
Species & Abundance \\
\hline
H$_2$ & 0.5 \\
HD & 1.00E-05 \\
He & 9.00E-02 \\
C & 1.40E-04 \\
N & 7.50E-05 \\
O & 3.20E-04 \\
Na & 2.25E-09 \\
Mg & 1.09E-08 \\
Si & 9.74E-09 \\
P & 2.16E-10 \\
S & 9.14E-08 \\
Cl & 1.00E-09 \\
Fe & 2.74E-09 \\
\hline
\end{tabular}\\
\end{minipage}
\end{table}
%
\begin{table*}
 \centering
 \begin{minipage}{160mm}
\caption{Summary of the energies characterizing the mobility of icy species on grains in the model.}
\label{tab-en}
  \begin{tabular}{@{}llcllc@{}}
  \hline	
\multicolumn{3}{c}{Surface} $|$& \multicolumn{3}{c}{Bulk ice (mantle sublayers)} \\
Energy & Notation & Value & Energy & Notation & Value \\
  \hline
Desorption (adsorption) & $E_D$ & . . . \footnote{Data from the network made available by \citet{Albertsson13}.} & Absorption & $E_B$ & 3.0$E_D$\footnote{\citet{Kalvans15a,Kalvans15c}} \\
Binding (diffusion) & $E_{b,s}$ & $0.35E_D$\footnote{\citet{Garrod11}} & Binding (diffusion) & $E_{b,m}$ & 0.35$E_B$ \\
  \hline
\end{tabular}\\
\end{minipage}
\end{table*}
Table\,\ref{tab-el} shows the initial abundances of chemical species. The adopted abundance of deuterium nuclei is $1.0\times10^{-5}$ relative to H nuclei, consistent with findings that the overall cosmic D/H is lower in regions with higher H column density and lower temperatures \citep{Linsky06,Prodanovic10}. Abundances for major elements were taken from \citet{Garrod08}, those for elements heavier than oxygen -- from \citet{Aikawa99}. This means that we consider a medium where refractory elements are significantly depleted, while the ice forming species are not. This approach is similar to that of \citet{Furuya15}.

Self-shielding of HD, and its mutual shielding by H$_2$ was added \citep{Lepetit02,Wolcott11}, in addition to the shielding of gaseous and adsorbed H$_2$, CO, and N$_2$. For consistency, we include self and mutual shielding for D$_2$ in the same manner as for HD. The latter aspect has little effect, because of the low abundance of D$_2$.

Neutral gaseous species can be adsorbed on to interstellar grains with radius $a+b$, where $a=0.1\, \mu$m is the radius of grain nucleus and $b$ is ice thickness. The latter is a time-dependent quantity, expressed as
   \begin{equation}
   \label{chem2}
b=\frac{N_{\mathrm{ice}}}{N_s}\times b_m,
   \end{equation}
where $N_{\rm ice}$ is the total number of ice molecules per grain, $N_s=1.5\times10^6$ is the number of adsorption sites on the grain surface and also per monolayer (ML), and $b_m=3.5\times10^{-8}$\,cm is the assumed monolayer thickness.

The ice mantle was described as consisting of four layers -- one surface and three subsurface layers (`sublayers'). Chemical reactions may occur in all sublayers, in line with the approach developed by \citet{Kalvans15b}. Dissociation of icy species by interstellar and cosmic-ray-induced photons is possible; it is assumed that subsurface species are shielded by the layers above. Each ML has a radiation attenuation probability of 0.007 \citep{Andersson08}.

We note that experiments show that the evaporation of icy mixtures occurs sequentially, i.e., through diffusion from subsurface ice layers \citep{Martin14}, although partial entrapment is possible \citep{Fayolle11}. This supports our model with mobile bulk-ice molecules. Chemical photoprocessing of such molecules is important because it can significantly influence the abundances of minor icy species, such as COMs \citep[e.g., ][]{Kalvans10,Garrod13}.

The rate of desorption of surface molecules is governed by their adsorption energies $E_D$. These correspond to bulk absorption energies of $E_B=3E_D$ for species in the sublayers of the icy mantle \citep{Kalvans15c}. Diffusion is governed by binding energies $E_{b,s}=xE_D$ and $E_{b,m}=xE_B$ for ice surface and mantle (sublayer) phases, respectively. The value of the parameter '$x$' was taken to be 0.35 \citep{Garrod11}. The various energies of the icy species are summarized in Table~\ref{tab-en}. The present study does not consider quantum tunneling for molecule diffusion on the surface and in the ice, which agrees with \citet{Katz99} and \citet{Watanabe10}. Inter-sublayer diffusion for ice molecules was included \citep{Kalvans15b}, as it is essential in transferring molecules from bulk ice to the surface when the ice temperature rises in Phase~2.

Six desorption mechanisms for species in the surface layer were considered -- desorption by interstellar photons, cosmic-ray-induced photodesorption, reactive desorption, indirect reactive desorption, evaporation, and desorption by cosmic-ray-induced whole-grain heating. The efficiency for indirect reactive desorption -- or desorption by H+H surface reaction heat -- was calibrated using the approach explained in detail in the previous works \citep{Kalvans15a,Kalvans15b}. Evaporation and photodesorption of species in subsurface layers was allowed if the number of the above MLs was lower than unity.

Because the molecular desorption energy $E_D$ is lower than the energy required for bulk diffusion, $E_{b,m}$, ice evaporation rate at rising temperatures is limited by diffusion, not thermal desorption. This is in an agreement with the experimental results by \citet{Oberg09} and \citet{Mispelaer13}. Such a behavior arises because in dense icy mixtures the (evaporating) surface becomes saturated with refractory species and the evaporation is hampered, unless the volatile molecules are able to diffuse efficiently from the bulk ice to the surface layer.

The rate coefficient (s$^{-1}$) for desorption by interstellar and secondary photons was calculated according to
   \begin{equation}
   \label{chem3}
k_{\mathrm{pd}}=\frac{\pi a^2 F_{\mathrm{ph}} Y_{\mathrm{ph}}}{N_s},
   \end{equation}
where $F_{\rm ph}$ (s$^{-1}$cm$^{-2}$) is photon flux ($1.7\times10^8$\,cm$^{-2}$\,s$^{-1}$ for interstellar photons at $A_{\rm V}=0$\,mag and 4875\,cm$^{-2}$\,s$^{-1}$ for cosmic-ray-induced photons). Photodesorption yield $Y_{\rm ph}$ was taken to be $3\times10^{-4}$ for interstellar photons \citep{Arasa15,Furuya15} and a 1.5 times lower value for cosmic-ray-induced photons \citep{Kalvans15b}. Eq.\,\ref{chem3} is physically more adequate than the approach of \citet{Kalvans15b} because the ratio between $N_s$ and grain cross section $\pi a^2$ is constant for spherical grains of different sizes. In other words, Eq.\,(\ref{chem3}) means that the actual photodesorption rate is independent of the grain size and ice thickness $b$. The calculated photodesorption rates are lower than in our previous studies. A standard cosmic-ray ionization rate of $1.3\times10^{-17}$\,s$^{-1}$ was used.

\section{Results}
\label{res}

In this section, we briefly mention the general processes governing ice chemistry. Then, we turn to the novelties of our deuterium chemistry model before focusing on the calculation results in the context of observations of circumstellar deuterated molecules. We remind that the simulation starts at $t=-1.5$\,Myr, while the protostar is formed at 0\,Myr, and the envelope is modeled up to 0.5\,Myr, at which point it reaches a maximum temperature of 200\,K. We try to attribute chemical processes to gas or solid (ice) phases although these phases are closely intertwined. For example, while the most efficient deuteration occurs in the gas-phase, depletion of heavy species and formation of H$_2$ on grain surfaces are prerequisites for this process.
%
\begin{figure*}
 \vspace{-1cm}
  \hspace{1cm}
  \includegraphics[width=18.0cm]{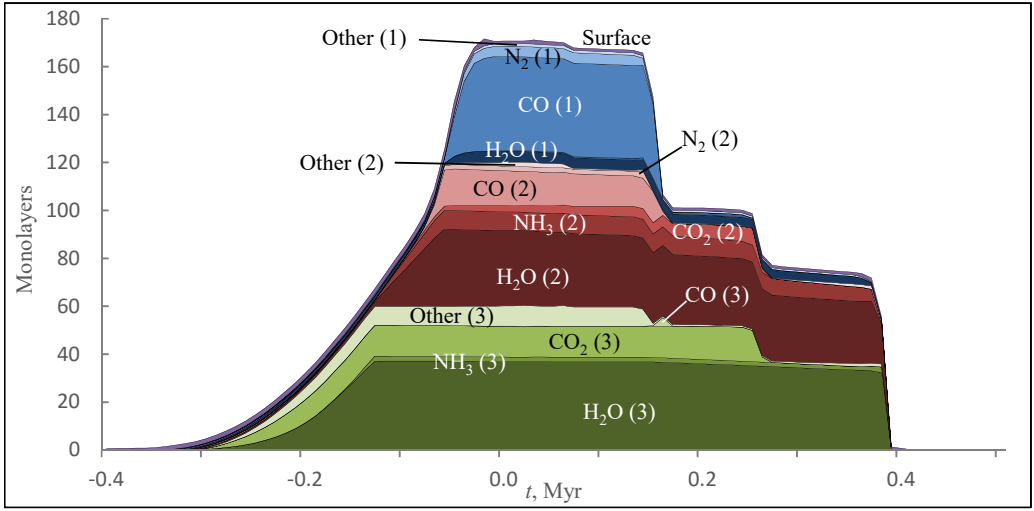}
 \vspace{-17cm}
 \caption{Calculated ice thickness and abundance in monolayers for major species in the sublayers. The numbers in parentheses indicate the number of the respective sublayer, while `Other' stands for all other icy species in that sublayer. Note that in the model, molecules within a sublayer are fully mixed.}
 \label{att-sub}
\end{figure*}
\subsection{Evolution of the ice layer}
\label{r-ice}

The first water ice monolayer is formed at $A_{\rm V}=3.0$\,mag ($t=-0.32$\,Myr). This is consistent with observations \citep[e.g.,][]{Whittet01} and some models \citep{Garrod11,Taquet14,Hocuk15}. Ice appears later than in the model of \citet{Furuya15}. This is largely because the simulation of \citeauthor{Furuya15} starts at an earlier point in cloud evolution and takes longer before the onset of rapid core collapse.

General ice chemistry and processes with similar model for star-forming cores have been described before \citep{Kalvans15a,Kalvans15b}. Fig.\,\ref{att-sub} shows the calculated abundances for major ice species in specific sublayers. The evolution of total ice thickness can also be seen. After the appearance of the first monolayer of adsorbed species on grain surface, the freeze-out proceeds at an increasing rate thanks to the growing $n_{\rm H}$ and decreasing $T_{\rm dust}$. Species with high $E_D$ (e.g., water and ammonia) become depleted at $t\approx-0.05$\,Myr. The depicted temporal evolution of the ice layer on grains provides the context for the discussion that follows. Five species -- H$_2$O, CO, CO$_2$, NH$_3$, and N$_2$ -- are major ice ingredients. Together they constitute 98 per cent of all ice molecules at $t=0$. In the model, CO and N$_2$ are products of gas-phase chemistry, while $>90$ per cent of H$_2$O, CO$_2$, and NH$_3$ molecules are formed via surface reactions.

Ice species are initially formed either in the gas phase, and then get adsorbed, or directly formed on the surface, before they are buried below ice surface. In total, about 10 per cent of molecules are chemically transformed in bulk ice. The subsurface (and other) reactions are driven by interstellar photons in the translucent cloud core (up to $A_{\rm V}\approx7$\,mag; $t\leq-0.16$) and cosmic-ray-induced photons in the dark core and the envelope. The subsequent heating of the protostellar envelope releases the icy species. Calculation results show that CO, CO$_2$, and H$_2$O evaporate at approximate temperatures of 24, 58, and $>$115\,K, respectively. Because the molecular diffusion energy in bulk ice $E_{b,m}$ is higher than desorption energy $E_D$, species often diffuse to the surface and evaporate over an extended period of time.

\subsection{Deuterium chemistry in ice}
\label{r-dice}
%
\begin{figure*}
 \vspace{-1.5cm}
  \hspace{-1.0cm}
  \includegraphics[width=18.0cm]{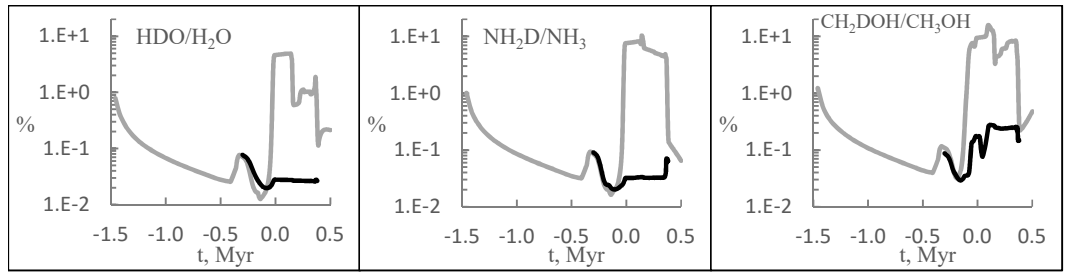}
 \vspace{-20cm}
 \caption{Calculated evolution of subsurface bulk-ice (black) and mantle surface (grey) D/H ratios for important hydrogenated species. The bulk ice contains $>90$ per cent of these molecules until evaporation.}
 \label{att-rd-ice}
\end{figure*}
%
%
\begin{figure}
 \vspace{-1.0cm}
  \includegraphics[width=18.0cm]{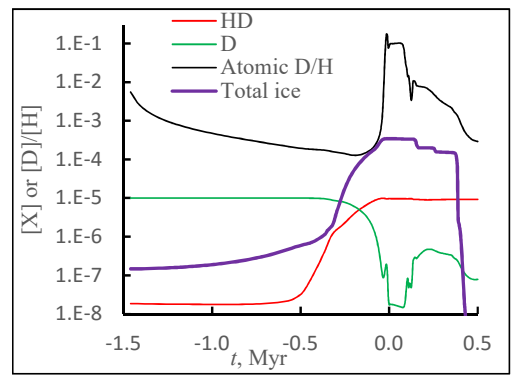}
 \vspace{-19cm}
 \caption{Calculated atomic D and HD abundances, and the gas-phase ratio for atomic D with respect to atomic H. The evolution of the total relative abundance of all icy species is also shown. One ice monolayer on the grain surface corresponds to a total ice species abundance of $2\times10^{-5}$.}
 \label{att-h}
\end{figure}
Fig.\,\ref{att-rd-ice} shows the calculated deuterium fractionation of the most important hydrogenated ice species for surface and subsurface mantle (sublayers 1, 2, and 3). The limited time-span for $R_{\rm D}$ curve of bulk ice species indicates the period of existence for the subsurface layers. The figure illustrates the concentration of deuterated species on the surface. The HDO/H$_2$O surface abundance ratio can reach 3 per cent, although average HDO/H$_2$O in ice does not exceed 0.2 per cent at any time when a considerable ice mass ($\geq1$\,ML) is present. This is in agreement with the non-detection of solid HDO in the interstellar or circumstellar medium. The upper limit for HDO$/$H$_2$O abundance ratio in ice is 2 per cent \citep{Parise03,Dartois03}. $R_{\rm D}$ for water ice (bulk) depends on the availability of atomic D for reactions on grain surfaces during ice accumulation in Phase~1.

Deuterium fractionation of \textit{surface} molecules in the translucent core is affected by two competing effects. First, photodissociation of HD and D$_2$ occurs at a higher rate than that of H$_2$, because of self-shielding effects. This ensures elevated atomic D/H, facilitating deuteration (especially visible in the early stages in Figs. 3 and 4; see also Sect.~\ref{nchem}). Second, species on the surface undergo repeated photodissociation and recombination (which occurs faster with H atoms), before being incorporated in the bulk ice. This results in a low molecular $R_{\rm D}$. Both effects were initially found by \citet{Furuya15}.

The dip in the ice deuteration curves, visible for all species depicted in Fig.\,\ref{att-rd-ice} at $t\approx-0.10$\,Myr, arises because the ice formation epoch partially overlaps with a rapid synthesis of molecular hydrogen (including HD) on grains. D atoms are consumed in surface reactions with atomic H faster than they are accreted from the gas phase. Because the majority of D atoms goes into HD, ice species formed during this period are poor in deuterium. This coincidence at $-0.10$\,Myr is illustrated with Fig.\,\ref{att-h}.

Afterwards, the freeze-out removes all heavy hydrogenated species from the gas. The abundance of H and D atoms is low because hydrogen has been locked in molecules. Immediately after the rapid accretion phase, conditions for efficient deuterium fractionation are maintained by gas-phase chemistry. These deuteration processes, however, affect only the outer surface of ices.

\subsection{Deuterium fractionation of gaseous species}
\label{rspec}
%
\begin{figure*}
 \vspace{-1.5cm}
  \includegraphics[width=18.0cm]{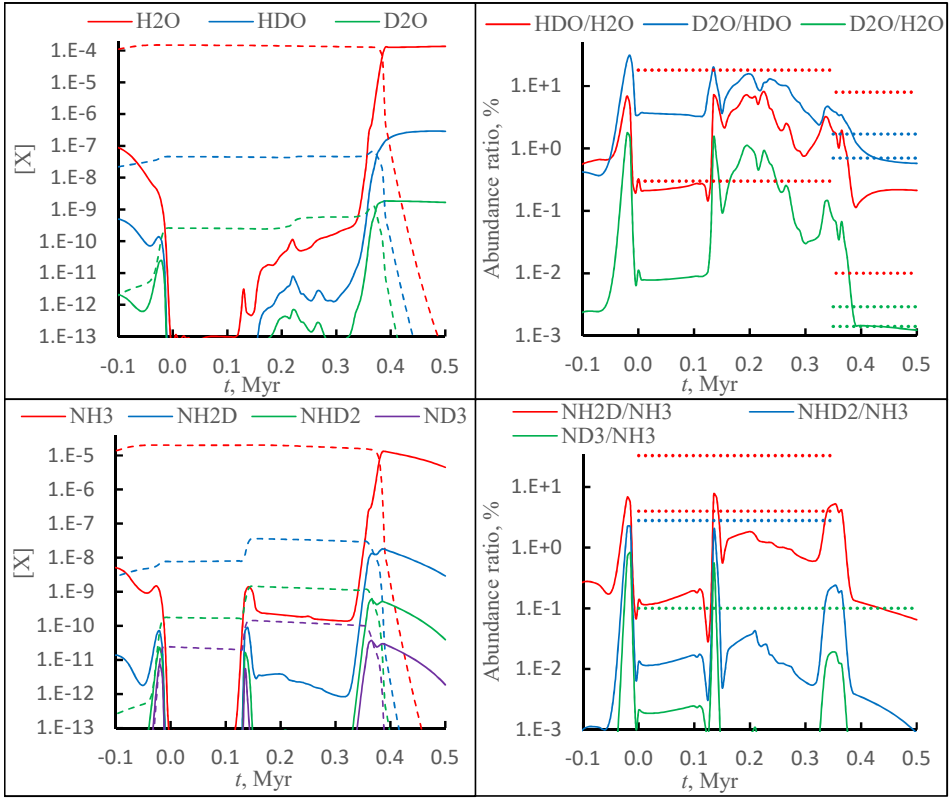}
 \vspace{-13cm}
 \caption{Evolution of calculated relative abundances [X] (left) and $R_{\rm D}$ ratios (right) for water and ammonia in the circumstellar envelope. Solid and dashed lines in the left-hand panels indicate gas and ice-phase species, respectively. Horizontal dotted lines in the right-hand panels indicate observational constraints (upper and lower $R_{\rm D}$ limits) from Table\,\ref{tab-obs}.}
 \label{att-g1}
\end{figure*}
%
\begin{figure*}
 \vspace{-1.5cm}
  \hspace{3cm}
  \includegraphics[width=18.0cm]{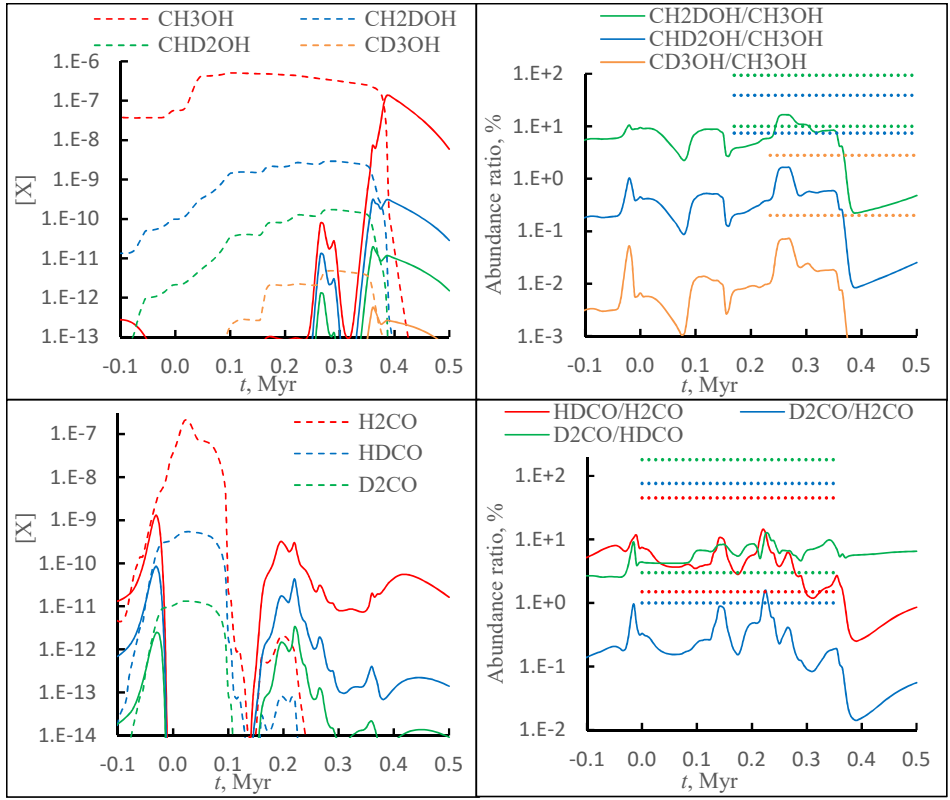}
 \vspace{-13.5cm}
 \caption{Evolution of calculated relative abundances (left) and $R_{\rm D}$ ratios (right) for methanol and formaldehyde as in Fig.\,\ref{att-g1}. The abundance of CH$_3$OD (not shown for clarity) is, on average, about half of that of CH$_2$DOH.}
 \label{att-g2}
\end{figure*}
%
\begin{table*}
 \centering
 \begin{minipage}{130mm}
\caption{Observational data on deuterium fractionation ratios for gas-phase chemical species in low-mass protostars and agreement with the model.}
\label{tab-obs}
  \begin{tabular}{@{}lcclc@{}}
  \hline
Isotopologues & Assumed $T$ & Abundance & References & R$_{\rm D}$ \\
 & range, K & ratio, per cent &  & agreement?\footnote{Agreement between observations and calculations. If no, the closest calculated value is indicated.} \\
 \hline
HDO/H$_2$O & $<$100 & 0.3--18 & (1, 2) & yes \\
HDO/H$_2$O & $>$100 & 0.01--8 & (2, 3, 4) & yes \\
D$_2$O/HDO & $>$100 & 0.7--1.7 & (4) & yes \\
D$_2$O/H$_2$O & $>$100 & 1.4E-3--2.9E-3 & (4) & yes \\
NH$_2$D/NH$_3$ & $<$100 & 4--33 & (5, 6, 7) & yes \\
NHD$_2$/NH$_3$ & $<$100 & 2.6--3.0 & (5) & 1.9 \\
ND$_3$/NH$_3$ & . . . & 0.1 & (8) & yes \\
 \hline
N$_2$D$^+$/N$_2$H$^+$ & $<$100 & 0.2--36 & (9, 10) & yes \\
DCO$^+$/HCO$^+$ & . . . & 0.39--14 & (5, 6, 11, 12, 13) & yes \\
DCN/HCN & $>$25 & 0.47--19.5 & (6, 11, 12, 13, 14) & yes \\
DNC/HNC & $>$25 & 1.5--6.1 & (12, 13) & yes \\
HDS/H$_2$S & $>$20 & 5--15 & (12) & yes \\
\hline
HDCO/H$_2$CO & $<$100 & 1.5--45 & (12, 13, 14, 15, 16) & yes \\
D$_2$CO/HDCO & $<$100 & 3--180 & (15, 16) & yes \\
D$_2$CO/H$_2$CO & $<$100 & 1--76 & (5, 10, 15, 16, 17) & yes \\
CH$_2$DOH/CH$_3$OH & $>$30 & 10--94 & (16, 18, 19) & 17\footnote{See Sect.\,\ref{rspec}.} \\
CH$_3$OD/CH$_3$OH & $>$30 & 0.8--6.9 & (16, 18, 19) & yes \\
CH$_3$OD/CH$_2$DOH & $>$30 & 4.8--11.4 & (16) & 26 \\
CHD$_2$OH/CH$_3$OH & $>$30 & 7--39 & (16, 18, 19) & 1.7 \\
CD$_3$OH/CH$_3$OH & $>$50 & 0.2--2.8 & (19) & 0.07 \\
DCOOCH$_3$/HCOOCH$_3$ & $>$100 & $\approx15$ & (20) & 4.6 \\
CH$_2$DOCH$_3$/CH$_3$OCH$_3$ & . . . & $\approx15$ & (21) & yes \\
\hline
C$_2$D/C$_2$H & . . . & 6--27 & (12, 13) & yes \\
C$_3$D/C$_3$H & . . . & 2.7--6.2 & (22) & yes \\
C$_3$HD/C$_3$H$_2$ & . . . & 4.8--9.4 & (22) & yes \\
C$_4$D/C$_4$H & . . . & 1.3--2.3 & (22) & yes \\
C$_4$HD/C$_4$H$_2$ & . . . & 1.3--5.1 & (22) & yes \\
DC$_3$N/HC$_3$N & . . . & 2.0--5.1 & (22, 23) & yes \\
DC$_5$N/HC$_5$N & . . . & 2.2--4.0 & (22) & yes \\
\hline
\end{tabular}\\
References: (1) \citet{Liu11,Persson13}; (2) \citet{Coutens13b}; (3) \citet{Taquet13b,Persson14}; (4) \citet{Coutens14a}; (5) \citet{Loinard01}; (6) \citet{Shah01}; (7) \citet{Hatchell03}; (8) \citet{Tak02}; (9) \citet{Emprechtinger09,Tobin13}; (10) \citet{Roberts07}; (11) \citet{Jorgensen04}; (12) \citet{Dishoeck95}; (13) \citet{Schoier02}; (14) \citet{Roberts02}; (15) \citet{Loinard00}; (16) \citet{Parise06}; (17) \citet{Ceccarelli01}; (18) \citet{Parise02}; (19) \citet{Parise04}; (20) \citet{Demyk10}; (21) \citet{Richard13}; (22) \citet{Sakai09b}; (23) \citet{Cordiner12}.
\vspace{-0.3cm}
\end{minipage}
\end{table*}
Chemical models tend to produce higher-than-observed maximum gas-phase abundances for major ice species, such as water, that are evaporating in the vicinity of a protostar \citep[e.g.][]{Garrod06,Taquet14,Wakelam14}. This discrepancy probably arises because the observations often have a limited resolution and sample a whole line-of-sight in the nebula. Despite this, the observed molecular ratios, such as $R_{\rm D}$, can be reliable. The necessary condition is that most of the molecules of the observed species occur in regions with similar history and physical conditions. This can be verified by accurately evaluating the excitation temperatures of the different isotopologues \citep{Nishimura13}. While calculated $R_{\rm D}$ values are our main result, we also present calculated abundances. This allows evaluating whether the $R_{\rm D}$ values are truly relevant.

We compare the numerical simulation results against observed gas-phase molecular $R_{\rm D}$ in low-mass protostars. It is assumed that the temporal evolution of calculated abundances may qualitatively represent a 1D spatial structure of the protostellar core \citep{Garrod08}. This means that higher temperatures in the warm-up stage of the model represent regions closer to the protostar.

The observational data on abundance ratios for isotopologues of different species have been summarized in Table\,\ref{tab-obs}. The calculated abundance and $R_{\rm D}$ curves are shown graphically in figures including those in Appendix~\ref{app-a} (available in electronic form). Observations with only upper limits reported were not considered here. For some species (e.g., water), the temperature ranges for Table\,\ref{tab-obs} were specified in the reference papers. For others, it was usually assumed that the excitation temperatures represent the lower limit of the kinetic temperature. Because the excitation temperatures can be very low, this condition does not always constrain the temperature range. The temperature range indicated for molecular $R_{\rm D}$ in Table\,\ref{tab-obs} is only authors' estimate to help classifying literature data for the purposes of the present paper.

Fig.\,\ref{att-g1} shows that for gas-phase water and ammonia, there are several peaks in the deuterium fractionation ratios. The first peak occurs a few kyr before starbirth, when the remaining gaseous heavy molecules are being completely depleted during freeze-out. Efficient deuterium enrichment occurs in the cold gas of the dense core \citep{Millar89,Roberts00,Roberts04}. While this peak corresponds to Phase~1 in our model, it might be relevant to the outer, colder parts of early circumstellar envelopes.

The second peak at 0.14\,Myr (22\,K) coincides with the evaporation of solid N$_2$ and CO. These species re-appear in the gas phase and are partially converted to N$_2$D$^+$ and DCO$^+$ by the products of cosmic-ray-induced dissociation and ionization of H$_2$ and He. N$_2$D$^+$ and DCO$^+$ then serve as deuteration agents. Meanwhile, the presence of abundant CO now removes the heavy isotopologues of H$_3^+$ that were the main deuteration agents in the cold phase. 

The third, smaller peak at 0.2\,Myr is caused by evaporation of methane and its D isotopologues. Significant amounts of methane ice are in inner ice layers, near the grain nuclei and have to diffuse to the surface before evaporating. The total evaporation of methane takes almost 20\,kyr. Methane can be transformed to CH$_3^+$ by radicals and ions in the gas. Most of these are generated by cosmic-ray induced dissociation and ionization of H$_2$ and He. CH$_3^+$ isotopologues are affected by reactions that are selective with respect to H and D atoms \citep[Table 13 of][]{Albertsson13}. The high deuteration of the abundant methane and its associated ions and radicals is transferred to other species via hydrogen atom exchanges or dispatchment of free H and D atoms. Because of these factors, the $R_{\rm D}$ peak, induced by methane evaporation, is about 50\,kyr long and ends only when most of CH$_4$ has been converted to CO.

This third $R_{\rm D}$ peak is followed by a number of smaller peaks caused by the successive evaporation of other molecules, such as C$_2$H$_2$, C$_2$H$_4$, H$_2$S, and CH$_3$OCH$_3$. The gas-phase processing of these species enable additional gas-phase deuterium fractionation reaction chains. The final gas-phase $R_{\rm D}$ peak at $t=0.34$\,Myr (94\,K) is related to the evaporation of water and ammonia themselves. The peak occurs before the bulk of these species has evaporated because deuterium-rich molecules are concentrated on the outer icy surface and evaporate first. After evaporation, $R_{\rm D}$ is mainly governed by gas-phase chemistry, much like the case of methane. With the evaporation of water, deuterated hydronium H$_3$O$^+$ replaces DCO$^+$ as the main molecular deuteration agent. Similar $R_{\rm D}$ peaks can be discerned for most of the other species. Appendix Fig.~\ref{att-p1} shows the calculated results for other (in)organic species observed in protostellar envelopes.

In the protostellar Phase~2, we obtain a higher gas-phase D$_2$O/HDO abundance ratio than HDO/H$_2$O for most of the time (cf. Fig.\,\ref{att-g1}), which agrees with the recent observations by \citet{Coutens14a}. We did not find similar results among recent papers on modeling of deuterium chemistry in circumstellar envelopes \citep{Aikawa12,Awad14,Taquet14}. The cause of the high D$_2$O/HDO abundance ratio is gas-phase chemistry -- this ratio is inherited from the high deuteration of the trihydrogen cation. Its four isotopologues H$_3^+$, H$_2$D$^+$, HD$_2^+$, and D$_3^+$ have almost equal relative abundances of $\approx10^{-11}$ at the beginning of Phase~2, thanks to the reactions proposed by \citet{Roberts04}. This deuterium enrichment is transferred to water and other molecules via ion-neutral reactions. Because a significant part of many species, including water, is formed on surfaces via the addition of atomic D, the $R_{\rm D}$ of these species never attains the high maximum values characteristic for purely gas-phase products, such as the formyl cation (DCO$^+$/HCO$^+=58$ per cent at 0.14\,Myr) or diazenylium (N$_2$D$^+$/N$_2$H$^+=81$ per cent).

Deuterated analogues of several COMs -- methanol CH$_3$OH, formaldehyde H$_2$CO, dimethyl ether CH$_3$OCH$_3$, and methyl formate HCOOCH$_3$ -- have been observed near low-mass protostars (Table\,\ref{tab-obs}). Fig.\,\ref{att-g2} shows the calculated abundances and $R_{\rm D}$ for methanol and formaldehyde deuterated isotopologues. Data for other COMs are visualized in Fig.~\ref{att-p2}. For CHD$_2$OH, CD$_3$OH, and DCOOCH$_3$ the high observed $R_{\rm D}$ values are not reached. In the case of CH$_2$DOH, the lowest $R_{\rm D}$ margin from observations is 10 per cent. This number is nominally reached in the model at $t=0.26$\,Myr and $T=60$\,K. At this temperature, methanol does not evaporate yet and its gas-phase relative abundance is below $10^{-10}$. This is orders of magnitudes lower than its maximum calculated abundance of $1.4\times10^{-7}$ during evaporation at 124\,K. At these later stages, $R_{\rm D}$ for CH$_2$DOH is only about 0.1 to 1 per cent. This discrepancy means that our model does not adequately reproduce the CH$_2$DOH/CH$_3$OH abundance ratio. A major reason for this disagreement is the synthesis of deuterium-poor methanol in bulk ice.

The synthesis of more than 90 per cent methanol CH$_3$OH, dimethyl ether CH$_3$OCH$_3$, and methyl formate HCOOCH$_3$ molecules occurs in the subsurface bulk ice layers \citep{Kalvans15b,Kalvans15c}. The latter two species have also gas-phase formation pathways that become effective when methanol and formaldehyde have evaporated \citep{Garrod06}. We use the low activation energies for CO hydrogenation from \citet{Kalvans15c}. This allows to reproduce the abundances of methanol and its daughter species. At the same time, formaldehyde H$_2$CO is largely consumed in ice before evaporation and is unable to reach the high relative abundance of about $1\times10^{-7}$ observed in protostellar envelopes \citep{Ceccarelli00,Ceccarelli01,Schoier02,Maret04}. Formaldehyde is produced in gas-phase to attain a relative abundance of nearly 10$^{-9}$ \citep[see][]{Garrod06}.

The deuterium fractionation of formaldehyde H$_2$CO is reproduced in our model, which has been a problem for some of the previous models \citep[e.g.][]{Aikawa12,Taquet14}. Similarly to the case of D$_2$O, the high efficiency of formaldehyde deuteration is thanks to to gas-phase chemistry with the network of \citet{Albertsson13}.

The deuteration of carbon-chain species is reproduced correctly (cf. Table\,\ref{tab-obs} and Fig.~\ref{att-p3}). The abundances and deuteration of some organic species become almost steady after the evaporation events, thanks to high-temperature ion-neutral reactions \citep[e.g.][]{Willacy09,Awad14}. In our model, carbon chains are formed mainly in reactions on grain outer surfaces. Exceptions are cyanides HC$_3$N and HC$_5$N (also inorganic species HCN and HNC) that can be formed in substantial amounts in subsurface ice layers rich in CO, N$_2$, or NH$_3$ \citep[Fig.~\ref{att-p1}; see also][]{Kalvans15c}.

\section{Conclusions}
\label{dconcl}

The scope and results of the simulation are broadly similar to those reported by \citet{Taquet14}. A major difference was that we consider bulk-ice chemical processes. It was found that \textit{subsurface photoprocessing facilitates the synthesis of deuterium-poor species in bulk ice}. Therefore, bulk ice chemistry does not help in explaining the very high observed $R_{\rm D}$ ratios for some COMs, contrary to a suggestion by \citet{Taquet14}. Solid deuterium-rich organic species are confined to ice surfaces and those abundant in the bulk ice have lower D/H.

With the cloud core contracting, darkening, and cooling in the prestellar Phase~1, atomic D is largely lost to synthesis of the HD molecule on grain surfaces. In our model, this process coincides with a period of rapid ice accumulation. Such competition for adsorbed D atoms results in that the majority of ice molecules are formed with relatively low $R_{\rm D}$.

The sequential evaporation of icy molecules in a warming protostellar envelope (Phase~2) affect the deuteration of many gaseous species. This happens either by the introduction of abundant deuteration agents in the gas (e.g., N$_2$D$^+$) or by providing abundant hydrogenated species (e.g., methane) that are processed by selective deuteration reactions to temporarily achieve high $R_{\rm D}$ level, which, in turn, is transferred to other species in reactions involving chemical radicals and ions. This results in species' $R_{\rm D}$ that is rapidly changing in the modeled temperature range of a protostellar envelope.

The evolution of molecular $R_{\rm D}$ with increasing temperature consists of a series of peaks.In the protostellar envelope, \textit{a combination of high gas-phase abundance and high D/H ratio is observed for species that are undergoing evaporation}. Additionally, $R_{\rm D}$ can be strongly affected by the evaporation of other species that take part in hydrogenation reaction chains (Sect.\,\ref{rspec}).

\section*{Acknowledgements}
We are grateful to the anonymous referees for many valuable comments and suggestions that made this paper better. JK, IS, and JRK thank Ventspils City Council for its persistent support. This research has made use of NASA's Astrophysics Data System.

 \footnotesize{
\bibliography{DDELAY24}
\bibliographystyle{mn2e}
 }

\label{lastpage}

\appendix
\onecolumn
\section{Calculated abundances and fractionation ratios for deuterated species observed around young stellar objects.}
\label{app-a}
 \vspace{5cm}

\begin{figure*}
 \vspace{-1cm}
  \includegraphics[width=21cm]{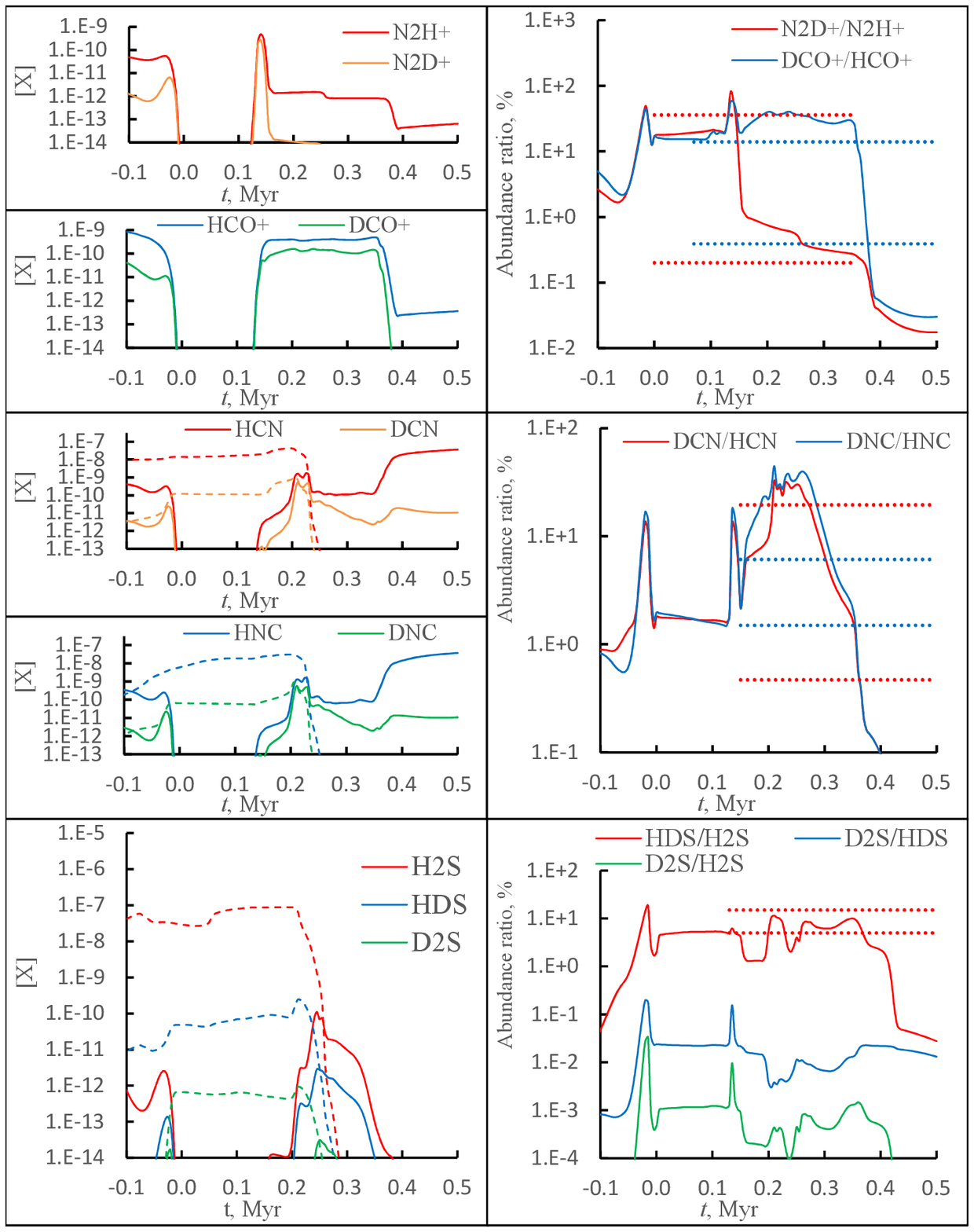}
 \vspace{-9.5cm}
 \caption{Evolution of calculated relative abundances [X] (left) and $R_{\rm D}$ ratios (right) for hydrogen isotopologues of observed gaseous (in)organic species in the circumstellar envelope. Solid and dashed lines in the left-hand panels indicate gas and ice-phase species, respectively. Horizontal dotted lines in the right-hand panels indicate observational constraints (upper and lower $R_{\rm D}$ limits) from Table\,\ref{tab-obs}.}
 \label{att-p1}
\end{figure*}
%
\begin{figure*}
 \vspace{-1.5cm}
  \hspace{3cm}
  \includegraphics[width=21.0cm]{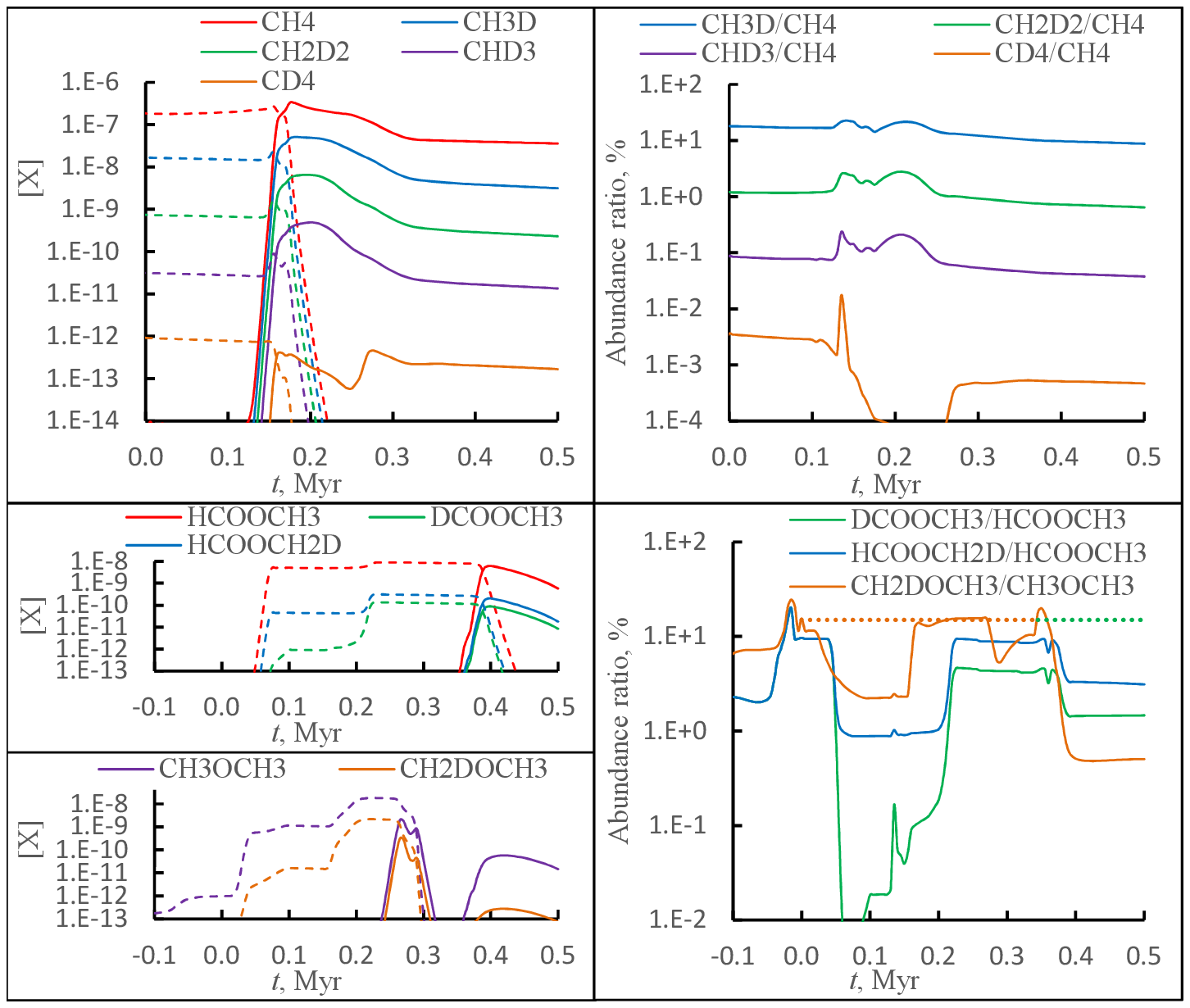}
 \vspace{-15.0cm}
 \caption{Evolution of calculated relative abundances (left) and $R_{\rm D}$ ratios (right) for methane, methyl formate, and dimethyl ether as in Fig.\,\ref{att-p1}.}
 \label{att-p2}
\end{figure*}
%
\begin{figure*}
 \vspace{-1.5cm}
  \hspace{2cm}
  \includegraphics[width=21.0cm]{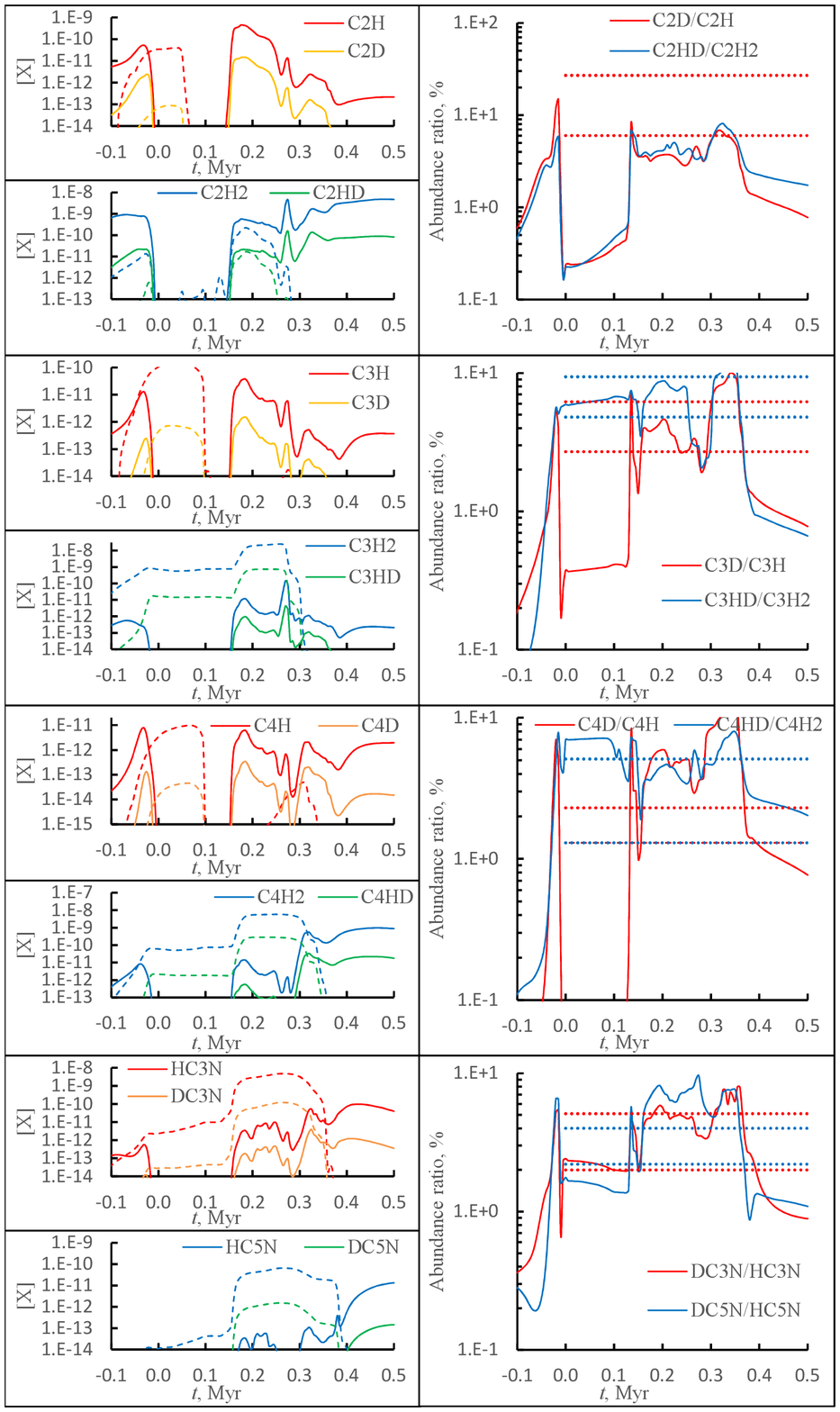}
 \vspace{-3.5cm}
 \caption{Evolution of calculated relative abundances (left) and $R_{\rm D}$ ratios (right) for observed carbon-chain species and acetylene as in Fig.\,\ref{att-p1}.}
 \label{att-p3}
\end{figure*}

\end{document}